\documentclass[twoside]{article}

\usepackage{times}
\usepackage{epsf}
 
\usepackage{fancyheadings,graphics}
\advance \headheight by 3.0truept       
 
\newlength{\nseparation}
\newenvironment{nfigure}[1]
        {\begin{figure}[#1]\hrule\vspace{\nseparation}\par}
        {\vspace{\nseparation}\par \hrule \end{figure}}

\usepackage{cite}   

\newcommand{\lt}{\left}
\newcommand{\rt}{\right}
\newcommand{\no}{\nonumber}
\newcommand{\nn}{\nonumber \\}

\newcommand{\ov}[1]{\overline{#1}}
\newcommand{\eq}[1]{(\ref{#1})}
\newcommand{\imag}{\mathrm{Im}\,}
 
\newcommand{\gev}{\,\mbox{GeV}}
\newcommand{\mev}{\,\mbox{MeV}}
\newcommand{\WA}{u}
\newcommand{\PI}{d}

\def\babar{\mbox{\sl B\hspace{-0.4em} {\scriptsize\sl      
A}\hspace{-0.4em} B\hspace {-0.4em} {\scriptsize\sl
      A\hspace{-0.1em}R}}}

\begin{document}

\title{{\normalsize FERMILAB-Conf-02/209-T\hfill hep-ph/0209008}\\[3cm]
Lifetimes of heavy hadrons\\[1mm] 
	beyond leading logarithms\footnote{Invited talk at 
\emph{Continuous Advances in QCD 2002/ARKADYFEST}\ (honoring the 60th 
birthday of Prof.~Arkady Vainshtein), 
17-23 May 2002, Minneapolis, Minnesota, USA.}\\[15mm]}

\author{Ulrich Nierste\\
Fermi National Accelerator Laboratory\\
Batavia, IL60510 -- 500, 
USA.\footnote{\uppercase{F}ermilab is operated by 
\uppercase{URA} under \uppercase{DOE} contract 
\uppercase{N}o.~\uppercase{DE-AC02-76CH03000}.
}\\ 
{\small E-mail: nierste@fnal.gov}}
\date{}


\maketitle

\begin{abstract}
The lifetime splitting between the $B^+$ and $B_d^0$ mesons has
recently been calculated in the next-to-leading order of QCD. These
corrections are necessary for a reliable theoretical prediction, in
particular for the meaningful use of hadronic matrix elements computed
with lattice QCD. Using results from quenched lattice
QCD we find $\tau(B^+)/\tau(B^0_d)=1.053\pm 0.016\pm 0.017$, where the
uncertainties from unquenching and $1/m_b$ corrections are not
included. The lifetime difference of heavy baryons $\Xi^0_b$ and
$\Xi^-_b$ is also discussed. 
\end{abstract}

\newpage
\thispagestyle{plain}
~
\newpage
\setcounter{page}{1}
\section{Introduction}
In my talk I present work done in collaboration with Martin Beneke,
Gerhard Buchalla, Christoph Greub and Alexander Lenz \cite{bbgln2}.

Twenty years ago the hosts of this conference showed that inclusive
decay rates of hadrons containg a heavy quark can be computed from
first principles of QCD. The \emph{Heavy Quark Expansion}\ (HQE)
technique \cite{hqe} exploits the heaviness of the bottom (or charm)
quark compared to the fundamental QCD scale $\Lambda_{QCD}$. In order
to study the lifetime of some $b$-flavored hadron $H$ containing a
single heavy quark one needs to compute its total decay rate
$\Gamma(H_b)$.  Now the HQE is an operator product expansion (OPE)
expressing $\Gamma(H_b)$ in terms of matrix elements of local $\Delta
B=0$ ($B$ denotes the bottom number) operators, leading to an
expansion of $\Gamma(H_b)$ in terms of $\Lambda_{QCD}/m_b$. In the
leading order of $\Lambda_{QCD}/m_b$ the decay rate of $H_b$ equals
the decay rate of a free $b$-quark, unaffected by the light degrees of
freedom of $H_b$. Consequently, the lifetimes of all $b$-flavored
hadrons are the same at this order.  The dominant source of lifetime
differences are weak interaction effects between the $b$-quark and the
light valence quark. They are depicted in Fig.~\ref{fig:lo} for the
case of the $B^+$--$B_d^0$ lifetime difference.
\begin{nfigure}{t!}
\centerline{
\epsfxsize=0.8\textwidth
\epsffile{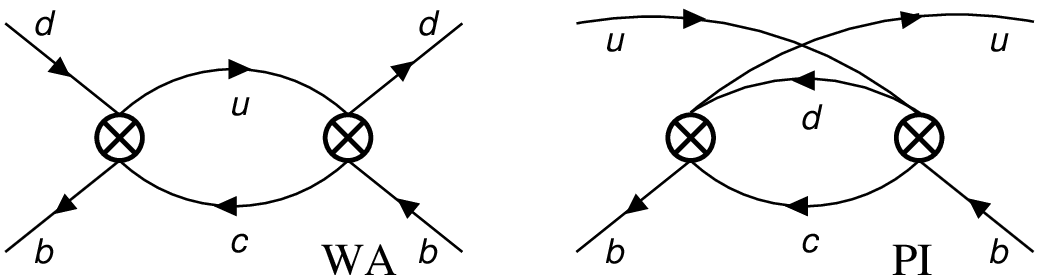}
}
\caption{\textit{Weak annihilation}\ (WA)
  and \textit{Pauli interference}\ (PI) diagrams in the leading order
  of QCD. They contribute to $\Gamma (B_d^0)$ and $\Gamma (B^+)$,
  respectively.  The crosses represent $|\Delta B|\!=\!1$ operators,
  which are generated by the exchange of $W$ bosons. CKM-suppressed
  contributions are not shown.}\label{fig:lo}
\end{nfigure}
The relative size of these weak non-spectator effects to the leading
free-quark decay is of order $16 \pi^2 (\Lambda_{QCD}/m_b)^3 ={\cal O}
(5\!\!-\!\!10\% )$.  The measurement of lifetime differences among
different $b$-flavored hadrons therefore tests the HQE formalism at
the third order in the expansion parameter.

The optical theorem relates the total decay rate $\Gamma (H_b)$ to the
self-energy of $H_b$:
\begin{eqnarray}
\Gamma (H_b) &=&
  \frac{1}{2 M_{H_b}} \langle H_b | {\cal T} | H_b \rangle .
  \label{opt}
\end{eqnarray}
Here we have introduced the transition operator:
\begin{eqnarray}
{\cal T} &=&  \imag i \! \int \!\! d^4 x \,
              T [H(x) \, H(0)]
                  \label{deft}
\end{eqnarray}
with the effective $|\Delta B|=1$ Hamiltonian $H$ describing the 
$W$-mediated decay of the $b$ quark. 
The HQE amounts to an OPE applied to ${\cal T}$ which
effectively integrates out the hard loop momenta (corresponding to the
momenta of the final state quarks). 
We decompose the result as 
\begin{eqnarray}
{\cal T} &=& \lt[ {\cal T}_0 \; + \; {\cal T}_2 \; + \; {\cal T}_3 \rt] 
              \lt[ 1 \; + \; 
                      {\cal O} ( 1/m_b^4 )
              \rt] \nn 
{\cal T}_3 &=& {\cal T}^{\WA} + {\cal T}^{\PI} + {\cal T}_{sing}.  
\label{t3}
\end{eqnarray}
Here ${\cal T}_n$ denotes the portion of ${\cal T}$ which is
suppressed by a factor of $1/m_b^n$ with respect to ${\cal T}_0$
describing the free quark decay. The contributions
to ${\cal T}_3$ from the weak interaction with the valence quark read
\begin{eqnarray}
{\cal T}^{\WA} &=& \frac{G_F^2 m_b^2 |V_{cb}|^2}{6 \pi} 
     \lt[\;\,\, |V_{ud}|^2 
     \lt(F^{\WA} Q^d \; + \; F_S^{\WA} Q_S^d \; + \;
          G^{\WA} T^d \; + \; G_S^{\WA} T_S^d \, \rt) \rt. \nn
&& \phantom{\frac{G_F^2 m_b^2 |V_{cb}|^2}{6 \pi} }
    \lt. +\, |V_{cd}|^2 
    \lt( \, F^c Q^d \,\, + \,\, F^c_S Q_S^d \; + \,\, G^c T^d \; + \;\,
 G^c_S T_S^d \, \rt)
     \rt] \nn 
&&  \phantom{\frac{G_F^2 m_b^2 |V_{cb}|^2}{6 \pi} } \, + \, (d \to s) \nn
{\cal T}^{\PI} &=& \frac{G_F^2 m_b^2 |V_{cb}|^2}{6 \pi} 
     \lt[ \,  F^{\PI} Q^u \; + \; F_S^{\PI} Q_S^u \; + \; 
          G^{\PI} T^u \; + \;  
          G_S^{\PI} T_S^u \, \rt] . \label{ope}  
\end{eqnarray}
Here $G_F$ is the Fermi constant, $m_b$ is the bottom mass and the
$V_{ij}$'s are elements of the Cabibbo-Kobayashi-Maskawa (CKM) matrix.
The superscript $q$ of the coefficients $F^q$, $F^q_S$, $G^q$, $G^q_S$
refers to the $cq$ intermediate state. The leading contributions to
${\cal T}^{\WA}$ and ${\cal T}^{\PI}$ are obtained from the left and
right diagram in Fig.~\ref{fig:lo}, respectively. They involve the
local dimension-6, $\Delta B=0$ operators
\begin{eqnarray}
Q^q   & = & \ov{b} \gamma_\mu (1-\gamma_5) q \, 
              \ov{q} \gamma^\mu (1-\gamma_5) b,\nn
Q_S^q & = & \ov{b} (1-\gamma_5) q \, \ov{q} (1+\gamma_5) b, \no\\[1mm] 
T^q   & = &   \ov{b} \gamma_\mu (1-\gamma_5) T^a q \, 
       \ov{q} \gamma^\mu (1-\gamma_5) T^a b, \nn
T_S^q & = & \ov{b} (1-\gamma_5) T^a q \, \ov{q} (1+\gamma_5) T^a b,
\label{ops}
\end{eqnarray}
where $T^a$ is the generator of color SU(3). The Wilson coefficients
$F^{\WA}\ldots G_S^{\PI}$ contain the physics from scales above $m_b$
and are computed in perturbation theory.  The remainder ${\cal
T}_{sing}$ in \eq{t3} involves additional dimension-6 operators, which
are $SU(3)_F$ singlets and do not contribute to the lifetime splitting
within the $(B^+,B_d^0)$ and $(\Xi_b^0,\Xi_b^-)$ iso-doublets. In order
to predict the widths $\Gamma (B_d^0)$ and $\Gamma (B^+)$ one needs to 
compute the hadronic matrix elements of the operators in \eq{ops}. 
After using the isospin relation $\langle B_d^0 | Q^{d,u} | B_d^0 \rangle =
\langle B^+ | Q^{u,d} | B^+ \rangle $ the matrix elements will enter 
$\Gamma (B_d^0) - \Gamma (B^+)$ in isospin-breaking
combinations, which are conventionally parametrized as \cite{ns,rome}
\begin{eqnarray}
\!\!\!\!\!\!\!\!\!&&
\langle B^+ | (Q^u - Q^d)  | B^+ \rangle \, = \,  
              f_B^2 M_B^2 B_1 ,\;\; 
\langle B^+ | (Q_S^u - Q_S^d)  | B^+ \rangle \, = \,  
              f_B^2 M_B^2 B_2 , \nn 
\!\!\!\!\!\!\!\!\!&&\langle B^+ | (T^u - T^d)  | B^+ \rangle \, = \,  
              f_B^2 M_B^2 \epsilon_1 ,\quad  
\langle B^+ | (T_S^u - T_S^d)  | B^+ \rangle \, = \,  
              f_B^2 M_B^2 \epsilon_2 . 
              \label{b} 
\end{eqnarray}
Here $f_B$ and $M_B$ are decay constant and mass of the $B$ meson,
respectively. In the
\emph{vacuum saturation approximation}\ (VSA) one has $B_1=1$,
$B_2=1+{\cal O}(\alpha_s(m_b),\Lambda_{QCD}/m_b)$ and
$\epsilon_{1,2}=0$.  Corrections to the VSA results are of order
$1/N_c$, where $N_c=3$ is the number of colors.

We now find from \eq{opt} and
\eq{ope}:
\begin{eqnarray}
\!\!\! \!\! \Gamma (B_d^0) - \Gamma (B^+) &  = & 
  \frac{G_F^2 m_b^2 |V_{cb}|^2}{12 \pi } \, f_B^2 M_B  
        \lt( |V_{ud}|^2 \vec{F}{}^{\WA} + |V_{cd}|^2 \vec{F}{}^{c} 
         - \vec{F}{}^{\PI} \rt) \! \cdot \! \vec{B} \!
    . \label{diffg}
\end{eqnarray}
Here we have introduced the shorthand notation
\begin{eqnarray}
 \vec{F}{}^q (z) \;  = \; 
\lt(
 \begin{array}{c}
   F^q (z) \\
   F_S^q (z) \\
   G^q (z) \\
   G_S^q (z) 
 \end{array} 
\rt) , && \quad
\vec{B} \;  = \; 
\lt(
\begin{array}{c}
  B_1  \\
  B_2  \\
  \epsilon_1 \\
  \epsilon_2 
\end{array}           
\rt) \qquad \mbox{for } q=\PI,\WA,c   . \label{short} 
\end{eqnarray}
Since the hard loops involve the charm quark, the coefficient
$\vec{F}{}^q$ depends on the ratio $z=m_c^2/m_b^2$.  The minimal way to
include QCD effects is the leading logarithmic approximation, which
includes corrections of order $\alpha_s^n \ln^n (m_b/M_W)$,
$n=0,1,\ldots$ in $\vec{F}{}^q$ in \eq{diffg}. The
corresponding leading order (LO) calculation of the width difference
in \eq{diffg} involves the diagrams in Fig.~\ref{fig:lo}
\cite{hqe,ns}.  Yet LO results are too crude for a precise calculation
of lifetime differences. The heavy-quark masses in \eq{diffg} cannot
be defined in a proper way and one faces a large dependence on
unphysical renormalization scales.  Furthermore, results for $B_{1,2}$
and $\epsilon_{1,2}$ from lattice gauge theory cannot be matched to
the continuum theory in a meaningful way at LO. Finally, as pointed
out in \cite{ns}, at LO the coefficients $F$, $F_S$ in \eq{diffg} are
anomalously small. They multiply the large matrix elements
parametrized by $B_{1,2}$, while the larger coefficients $G$, $G_S$
come with the small hadronic parameters $\epsilon_{1,2}$, rendering
the LO prediction highly unstable.  To cure these problems one must
include the next-to-leading-order (NLO) QCD corrections of order
$\alpha_s^{n+1}
\ln^n (m_b/M_W)$.  

The first calculation of a lifetime difference beyond the LO was 
performed for the $B_s^0$--$B_d^0$ lifetime
difference \cite{kn}, where ${\cal O}(\alpha_s)$ corrections were
calculated in the SU(3)$_{\rm F}$ limit neglecting certain terms of order
$z$. In this limit only a few penguin effects play a role. A complete
NLO computation has been carried out for the lifetime difference
between the two mass eigenstates of the $B_s^0$ meson in \cite{bbgln}.
In particular the correct treatment of infrared effects, which appear
at intermediate steps of the calculation, has been worked out in
\cite{bbgln}.  The recent computation in \cite{bbgln2} is conceptually
similar to the one in \cite{bbgln}, except that the considered
transition is $\Delta B=0$ rather than $\Delta B=2$ and the quark
masses in the final state are different. The NLO calculation of
$\Gamma (B_d^0) - \Gamma (B^+)$ involves the diagrams of
Fig.~\ref{fig:nlo}. In \cite{rome} the NLO corrections to $\Gamma
(B_d^0) - \Gamma (B^+)$ have been calculated for the limiting case
$z=0$. The corrections to this limit are of order $z\ln z$ or roughly
20\%. The first NLO calculation with the complete $z$ dependence was
presented in \cite{bbgln2} and subsequently confirmed in \cite{rome2}.
\begin{nfigure}{t!}
\centerline{
\epsfxsize=\textwidth
\epsffile{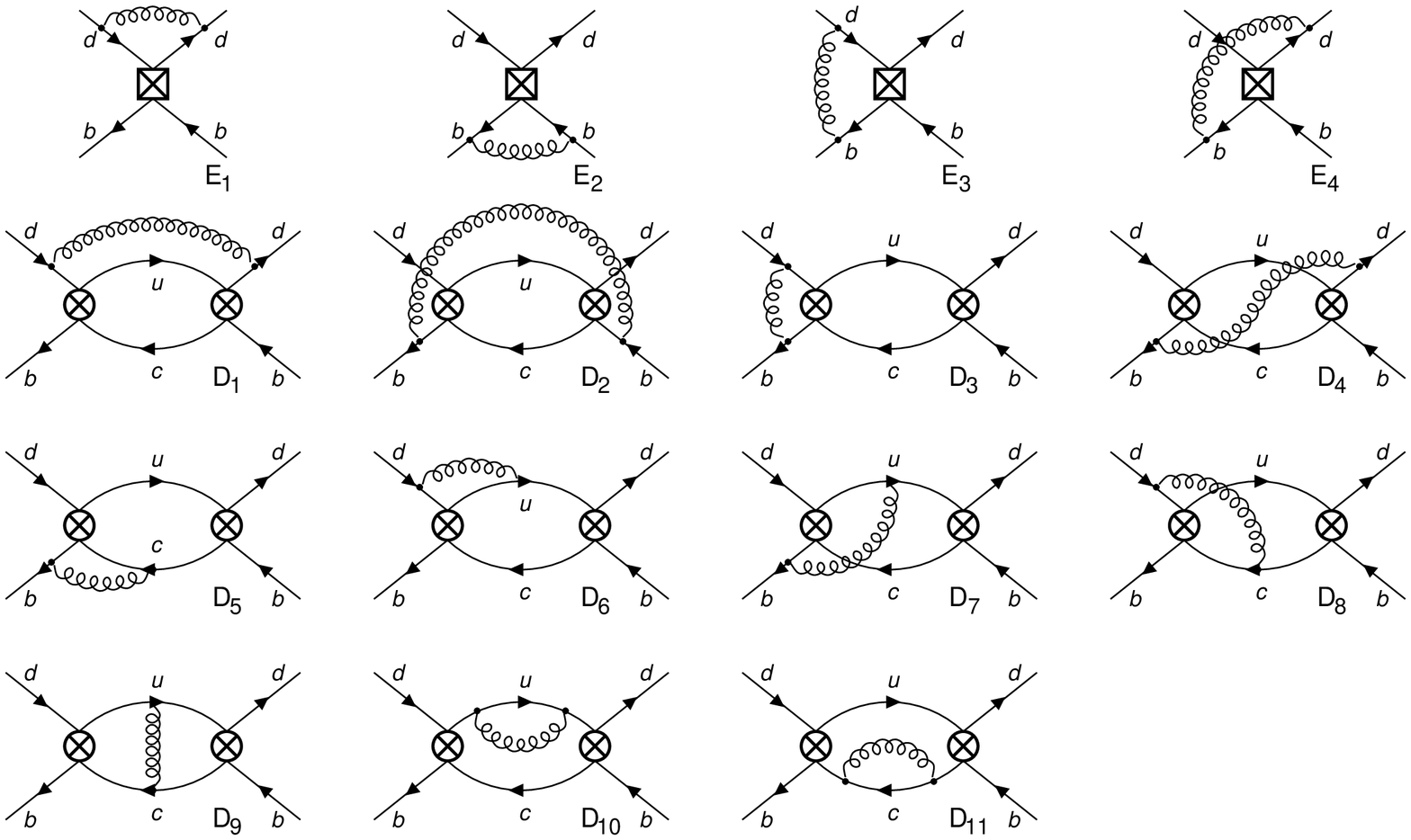}
}
\caption{WA contributions in the next-to-leading order of QCD. The PI
diagrams are obtained by interchanging $u$ and $d$ and reversing the 
fermion flow of the $u$ and $d$ lines. The first line shows the 
radiative corrections to $\Delta B\!=\!0$ operators, which are
necessary for the proper infrared factorization. 
Not displayed are the diagrams ${\rm E}_3^\prime$, ${\rm E}_4^\prime$
and ${\rm D}_{3-8}^\prime$ which are obtained from the corresponding 
unprimed diagrams by left-right reflection and the reverse of the
fermion flow.}\label{fig:nlo}
\end{nfigure}

\section{Lifetime differences at next-to-leading order}\label{sect:nlo} 

The analytic \mbox{expressions} for the Wil\-son co\-ef\-fi\-cients 
$F^{\WA,(1)}_{ij}- F^{\PI,(1)}_{ij} \ldots G^{\WA,(1)}_{S,ij} - G^{\PI
,(1)}_{S,ij}$ are cumbersome functions of $z$ involving
dilogarithms. They depend on the renormalization scheme chosen for the
$\Delta B\!=\!0$ operators in \eq{ops} and also on the renormalization
scale $\mu_0={\cal O}(m_b)$ at which these operators are
defined. These dependences properly cancel between $\vec{F}{}^q$ and
$\vec{B}$ in physical observables like \eq{diffg}. When our results
for $F^{\WA,(1)}_{ij} - F^{\PI,(1)}_{ij} \ldots G^{\WA,(1)}_{S,ij} -
G^{\PI ,(1)}_{S,ij}$ are combined with some non-perturbative
computation of $B_1,\ldots
\epsilon_2$, one has to make sure that the numerical values of these
hadronic parameters correspond to the same renormalization scheme. Our
scheme is defined by the use of dimensional regularization with
$\ov{\rm MS}$ \cite{bbdm} subtraction, an anticommuting $\gamma_5$ and
a choice of evanescent operators preserving Fierz invariance at the
loop level \cite{hn}. Choosing further $\mu_0=m_b$ the desired
lifetime ratio can be compactly written as
\begin{eqnarray} \lefteqn{
  \frac{\tau(B^+)}{\tau(B_d^0)} - 1 \; = \;  \tau(B^+) \,
                       \lt[ \Gamma (B_d^0) - \Gamma (B^+) \rt]} \nn & = &
0.0325 \, \lt( \frac{|V_{cb}|}{0.04} \rt)^2
        \, \lt( \frac{m_b}{4.8\gev} \rt)^2 \,
           \lt( \frac{f_B}{200\mev} \rt)^2 \,
        \times \nn && \!\!\!\!
      \Big[ \, ( 1.0 \pm 0.2) \, B_1 \; + \; (0.1 \pm 0.1) \, B_2 \; - \;
         (18.4 \pm 0.9) \, \epsilon_1 \; + \; (4.0 \pm 0.2) \, \epsilon_2
      \, \Big] .~~~ \label{phent}
\end{eqnarray}
Here $\tau(B^+) = 1.653\,$ps${}$ has been used in the overall
factor. 

The hadronic parameters have been computed in \cite{b} with quenched
lattice QCD using the same renormalization scheme as in the present
paper.  They read 
\begin{eqnarray}
\!\!\!\!\!\!\!\!
(B_1,B_2, \epsilon_1,\epsilon_2) \! &=& \!
 (1.10\pm 0.20, \, 0.79 \pm 0.10, \, -0.02 \pm 0.02, \, 0.03 \pm
 0.01 ) . 
\label{bec}
\end{eqnarray}
Inserting $|V_{cb}|=0.040 \pm 0.0016 $ from a CLEO analysis of inclusive
semileptonic $B$ decays \cite{cleo}, the world average $f_B = (200 \pm
30)\mev $ from lattice calculations \cite{r} and $m_b=4.8 \pm 0.1
\gev$ for the one-loop bottom pole mass into \eq{phent}, our NLO
prediction reads
\begin{eqnarray}
\frac{\tau(B^+)}{\tau(B_d^0)} &=& 1.053 \pm 0.016 \pm 0.017 
\label{res}
\end{eqnarray}
compared to 
\begin{eqnarray}
\lt[ \frac{\tau(B^+)}{\tau(B_d^0)} \rt]_{\rm LO} \; =\; 1.041 \pm 0.040 
\pm 0.013. \label{reslo}
\end{eqnarray}
Here the first error is due to the errors on the coefficients and
the hadronic parameters (\ref{bec}), and the second error is the
overall normalization uncertainty due to $m_b$, $|V_{cb}|$ and $f_B$
in (\ref{phent}). The Wilson coefficients also depend on the
renormalization scale $\mu_1$ at which the $\Delta B\!=\!1$ operators
entering the diagrams in Figs.~\ref{fig:lo} and \ref{fig:nlo} are
defined. This dependence stems from the truncation of the perturbation
series and diminishes order-by-order in $\alpha_s$. The dependence 
on $\mu_1$ is the dominant uncertainty of the LO prediction of the
lifetime ratio. In Fig.~\ref{fig:ldplot} the $\mu_1$-dependence of the
LO and NLO predictions for $\tau(B^+)/\tau(B^0_d)-1$ is shown. 
\begin{nfigure}{t!}
\centerline{
\epsfxsize=0.95\textwidth
\epsffile{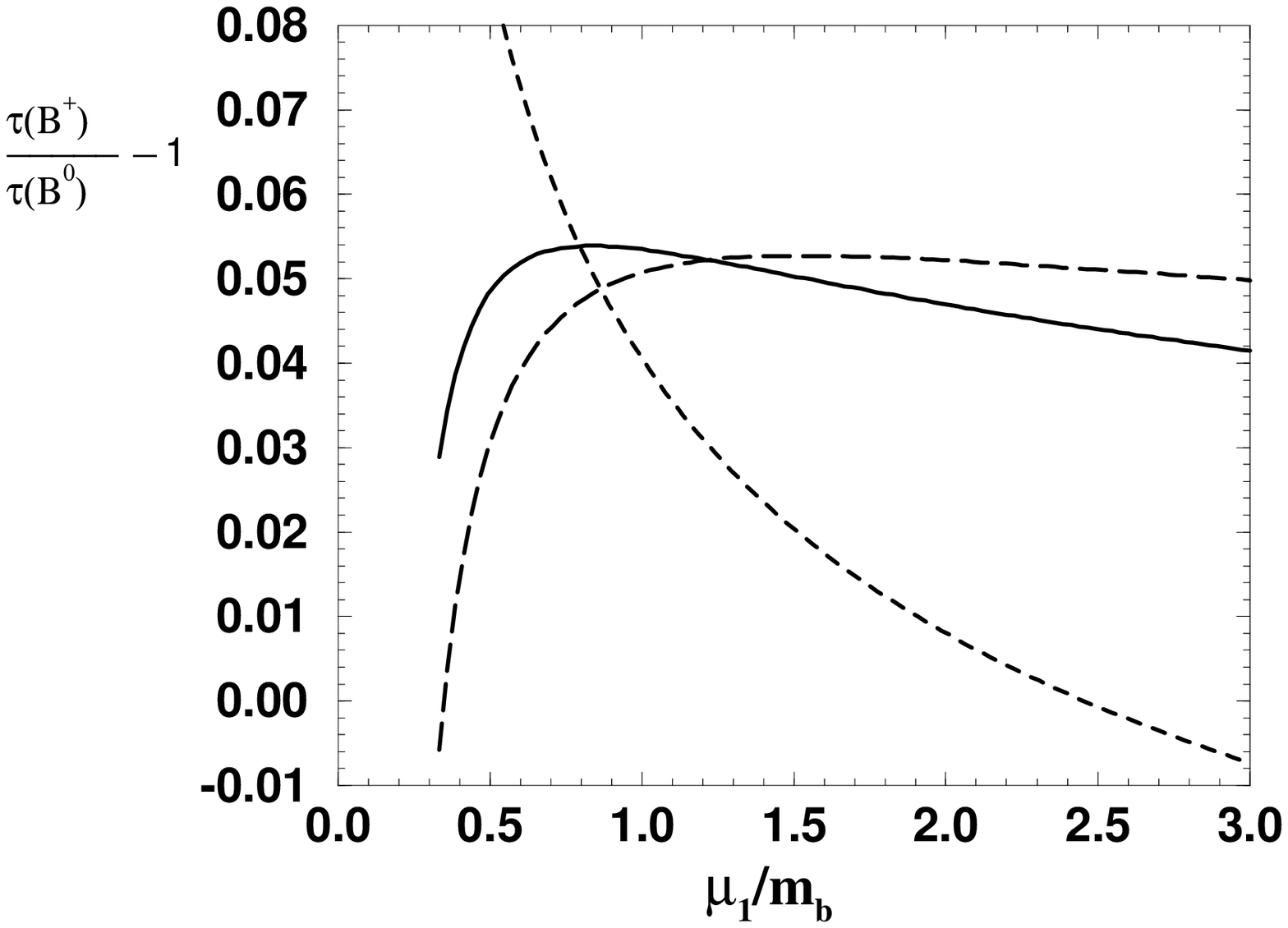}
}
\caption{Dependence of $\tau(B^+)/\tau(B^0_d)-1$
  on $\mu_1/m_b$ for the central values of the input parameters and
  $\mu_0=m_b$.  The solid (short-dashed) line shows the NLO (LO)
  result. The long-dashed line shows the NLO result in the
  approximation of \cite{rome}, i.e.\ $z$ is set to zero in the NLO
  corrections.  }\label{fig:ldplot}
\end{nfigure}
The substantial reduction of scale dependence at NLO leads to the
improvement in the NLO vs.\ LO results in \eq{res},\eq{reslo}.  
Note that the NLO
calculation has firmly established that $\tau(B^+)>\tau(B_d^0)$, a
conclusion which could not be drawn from the old LO result. The result
in \eq{res} is compatible with recent measurement from the B
factories \cite{babar,belle}:
\begin{eqnarray}
\frac{\tau(B^+)}{\tau(B_d^0)} &=& 
\left\{ \begin{array}{ll} 1.082 \pm 0.026 \pm 0.012 & {\rm (\babar)} \\ 
                          1.091 \pm 0.023 \pm 0.014 & {\rm (BELLE)}
        \end{array}\right.\no
\end{eqnarray}

The calculated Wilson coefficients can also be 
used to predict the  lifetime splitting within the iso-doublet
$(\Xi_b^0\sim bus,\Xi_b^- \sim bds)$ with NLO precision. 
The corresponding LO diagrams
are shown in Fig.~\ref{fig:bar}.
Note that the role of ${\cal T}^{\WA}$ and ${\cal T}^{\PI}$ is
interchanged compared to the meson case with ${\cal T}^{\WA}$
describing the Pauli interference effect. The lifetime difference
between $\Lambda_b \sim bud$ and $\Xi_b^0$ is expected to be small,
as in the case of $B_s^0$ and $B_d^0$, because it mainly stems from
the small U-spin breaking effects in the matrix elements appearing at
order $1/m_b^2$. 
\begin{nfigure}{t}
\centerline{
\epsfxsize=0.8\textwidth
\epsffile{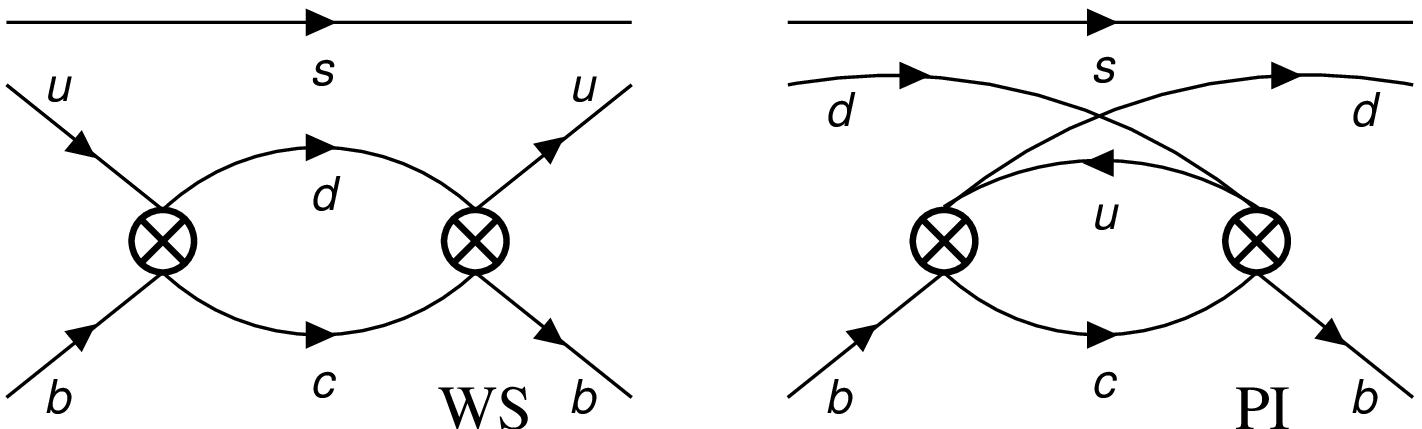}
}
\caption{\textit{Weak scattering}\ (WS) 
  and PI diagrams for $\Xi_b$ baryons in the leading order of QCD. They
  contribute to $\Gamma (\Xi_b^0)$ and $\Gamma (\Xi_b^-)$, respectively.
  CKM-suppressed contributions are not shown.  }\label{fig:bar}
\end{nfigure}

For $\Xi_b$'s the weak decay of the valence $s$-quark could be
relevant: the decays $\Xi_b^- \to \Lambda_b \pi^-$, $\Xi_b^- \to
\Lambda_b e^- \ov{\nu}_e$ and $\Xi_b^0 \to \Lambda_b \pi^0$ are
triggered by $s\to u$ transitions and could affect the total rates at
the ${\cal O}(1\%)$ level \cite{v}. Once the lifetime measurements
reach this accuracy, one should correct for this effect.
To this end we define
\begin{eqnarray}
\ov{\Gamma} (\Xi_b) \! &\equiv& \! 
  \Gamma (\Xi_b) - \Gamma (\Xi_b \to \Lambda_b X)
  \; = \; \frac{1- B(\Xi_b \to \Lambda_b X)}{\tau(\Xi_b)} 
  \; \equiv \; \frac{1}{\ov{\tau}(\Xi_b)}\nn
&&  \qquad\qquad\qquad\qquad\qquad\qquad\qquad\qquad
  \mbox{for }\, \Xi_b=\Xi_b^0,\Xi_b^-, \qquad
\end{eqnarray}
where $B(\Xi_b \to \Lambda_b X)$ is the branching ratio of the
above-mentioned decay modes. Thus $\ov{\Gamma} (\Xi_b)$ is the
contribution from $b\to c$ transitions to the total decay rate.
In contrast to the $B$ meson system, the matrix elements of the four
operators in \eq{ops} are not independent at the considered order in
$\Lambda_{QCD}/m_b$. Since the light degrees of freedom are in a
spin-0 state, the matrix elements $\langle \Xi_b | 2Q_S^q+Q^q | \Xi_b
\rangle$ and $\langle \Xi_b | 2T_S^q+T^q | \Xi_b \rangle$ are
power-suppressed compared to those in \eq{l} (see e.g.\ 
\cite{hqe,ns}). This, however, is not true in all renormalization
schemes, in the $\ov{\rm MS}$ scheme used by us $2Q_S^q+Q^q$ and
$2T_S^q+T^q$ receive short-distance corrections, because hard gluons
can resolve the heavy $b$-quark mass. A priori one can choose the
renormalization of e.g.\ $Q_S^q$ independently from $Q^q$, so that
$\langle \Xi_b | 2 Q_S^q+Q^q | \Xi_b\rangle={\cal O}
(\Lambda_{QCD}/m_b)$ can only hold in certain renormalization
schemes. This is also the case, if the operators are defined in heavy
quark effective theory (HQET) rather than in full QCD. After properly
taking into account these short-distance corrections, one can express
the desired lifetime ratio solely in terms of two hadronic parameters
defined as
\begin{eqnarray}
\langle \Xi_b^0 | (Q^u - Q^d) (\mu_0) | \Xi_b^0 \rangle 
& = & f_B^2 M_B M_{\Xi_b} \, L_1^{\Xi_b} (\mu_0), \nn
\langle \Xi_b^0 | (T^u - T^d) (\mu_0) | \Xi_b^0 \rangle 
& = & f_B^2 M_B M_{\Xi_b} \, L_2^{\Xi_b} (\mu_0)
 . \label{l} 
\end{eqnarray}
Then one finds
\begin{eqnarray}
  \frac{\ov{\tau}(\Xi_b^0)}{\ov{\tau}(\Xi_b^-)} - 1 & = &
    \ov{\tau}(\Xi_b^0) \, 
                       \lt[ \Gamma (\Xi_b^-) - \Gamma (\Xi_b^0) \rt] \nn
& = & 0.59   \, \lt( \frac{|V_{cb}|}{0.04} \rt)^2  
        \, \lt( \frac{m_b}{4.8\gev} \rt)^2 \, 
           \lt( \frac{f_B}{200\mev} \rt)^2 \, 
      \frac{\ov{\tau}(\Xi_b^0)}{1.5\, {\rm ps}} 
        \times \nn 
&& \qquad \qquad \qquad
      \Big[ \, ( 0.04 \pm 0.01) \, L_1  \; - \; 
               ( 1.00 \pm 0.04) \, L_2 \, \Big] 
               ,~~~ \label{phentxi}
\end{eqnarray}
with $L_i=L_i^{\Xi_b}(\mu_0=m_b)$. For the baryon case there is no
reason to expect the color-octet matrix element to be much smaller
than the color-singlet ones, so that the term with $L_2$ will 
dominate the result. The hadronic parameters $L_{1,2}$ have been
analyzed in an exploratory study of lattice HQET \cite{dsm} for
$\Lambda_b$ baryons. Up to SU(3)$_{\rm F}$ corrections, which are
irrelevant in view of the other uncertainties, $L_i^{\Xi_b}$ and
$L_i^{\Lambda_b}$ are equal.

\section{Conclusions}
Twenty years ago the ITEP group has developed the Heavy Quark
Expansion, which allows to study inclusive decay rates of heavy
hadrons in a model-free, QCD-based framework \cite{hqe}. The HQE
expresses these decay rates as a series in both $\Lambda_{QCD}/m_b$
and $\alpha_s(m_b)$.  With the advent of precision measurements of
lifetimes of $b$-flavored hadrons at the B factories and the Tevatron
correspondingly precise theory predictions are desirable. This
requires the calculation of higher-order terms in the HQE. The
inclusion of the $\alpha_s$ corrections presented in this talk is in
particular mandatory for any meaningful use of hadronic matrix
elements computed in lattice gauge theory. The calculated QCD
corrections to the WA and PI diagrams in
Figs.~\ref{fig:lo},\ref{fig:nlo} allow to study the lifetime splitting
within the $(B^+,B_d^0)$ and $(\Xi_b^0,\Xi_b^-)$ iso-doublets with NLO
accuracy. It is gratifying that these corrections have been
independently calulated by two groups finding agreement in their
analytic expressions for the Wilson coefficients \cite{bbgln2,rome2}.

Current lattice calculations, which are still in a relatively early
stage in this case, yield, when combined with our calculations,
$\tau(B^+)/\tau(B^0_d)=1.053\pm 0.016 \pm 0.017$ [see
(\ref{res})]. The effects of unquenching and $1/m_b$ corrections are
not included in the error estimate, but the unquenching effects can
well be sizable. A substantial improvement of the NLO calculation is
the large reduction of perturbative uncertainty reflected in the scale
dependence stemming from the $\Delta B=1$ operators.  This scale
dependence had been found to be very large at leading order,
preventing even an unambiguous prediction of the sign of
$\tau(B^+)/\tau(B^0_d)-1$ up to now \cite{ns}.

At present the experimentally measured $\Lambda_b$ lifetime falls
short of $\tau(B_d^0)$ by roughly 20\% \cite{pdg}, which has raised
concerns about the applicability of the HQE to baryons. Unfortunately
this interesting topic cannot yet be addressed at the NLO level for
two reasons: First, $\tau(\Lambda_b)/\tau(B_d^0)$ receives
contributions from the yet uncalculated SU(3)$_{\rm F}$-singlet
portion ${\cal T}_{sing}$ of the transition operator in
\eq{t3}. Second, the hadronic matrix elements entering
$\tau(\Lambda_b)/\tau(B_d^0)$ involve penguin (also called `eye')
contractions of the operators in \eq{ops}, which are difficult to
compute. These penguin contractions are contributions to the matrix
elements in which the light $q$ and $\ov{q}$ quark fields of the
operator are contracted with each other, not with the hadron's
valence quarks.

\section*{Acknowledgments}
I thank the organizers for the invitation to this wonderful
Arkadyfest workshop. I have enjoyed stimulating discussions with
many participants and look forward to future Arkadyfests, possibly 
on the occasions of Arkady's 70th, 80th, 90th and 100th birthdays! 
I am grateful to Martin Gorbahn for proofreading the manuscript.

\end{document}